# External beam irradiation angle measurement using a hybrid Cerenkov-scintillation detector

Emilie Jean[1,2,3], Simon Lambert-Girard[4], François Therriault-Proulx[4] and Luc Beaulieu[1,2]

[1] Département de physique, de génie physique et d'optique et Centre de recherche sur le cancer, Université Laval, Quebec, QC, Canada
[2] Département de radio-oncologie et Axe Oncologie du CRCHU de Québec, CHU de Québec - Université Laval, Quebec, QC, Canada
[3] Département de radio-oncologie du CIUSSS-MCQ, CHAUR de Trois-Rivières, Trois-Rivières, QC, Canada
[4] Medscint inc. Quebec, QC, Canada
E-mail: Luc.Beaulieu@phy.ulaval.ca

**Abstract**

Objective: In this study, we propose a novel approach designed to take advantage of the Cerenkov light angular dependency to perform a direct measurement of an external beam irradiation angle.
Approach: A Cerenkov probe composed of a 10-mm long filtered sensitive volume of clear PMMA optical fibre was built. Both filtered and raw Cerenkov signals from the transport fibre were collected through a single 1-mm diameter transport fibre. An independent plastic scintillation detector composed of 10-mm BCF12 scintillating fibre was also used for simultaneous dose measurements. A first series of measurements aimed at validating the ability to account for the Cerenkov electron energy spectrum dependency by simultaneously measuring the deposited dose, thus isolating signal variations resulting from the angular dependency. Angular calibration curve for fixed dose irradiations and incident angle measurements using electron and photon beams where also achieved.
Main results: The beam nominal energy was found to have a significant impact on the shapes of the angular calibration curves. This can be linked to the electron energy spectrum dependency of the Cerenkov emission cone. Irradiation angle measurements exhibit an absolute mean error of 1.86° and 1.02° at 6 and 18 MV, respectively. Similar results were obtained with electron beams and the absolute mean error reaches 1.97°, 1.66°, 1.45° and 0.95° at 9, 12, 16 and 20 MeV, respectively. Reducing the numerical aperture of the Cerenkov probe leads to an increased angular dependency for the lowest energy while no major changes were observed at higher energy. This allowed irradiation angle measurements at 6 MeV with a mean absolute error of 4.82°.
Significance: The detector offers promising perspectives as a potential tool for future quality assurance applications in radiotherapy, especially for stereotactic radiosurgery (SRS), magnetic resonance image-guided radiotherapy (MRgRT) and brachytherapy applications.

## 1. Introduction

Improvement of radiotherapy techniques has emphasized the need to perform dosimetric quality assurance tests to ensure that radiation dose is safely and correctly delivered to the tumour (Solberg et al 2012). Techniques such as stereotactic radiotherapy (SBRT) and intensity modulated radiotherapy (IMRT) employ a larger number of small fields and modulated beams (Meyer 2011, Elith *et al* 2011) while development of magnetic resonance image-guided accelerator (MR-LINAC) implies the presence of a magnetic field (Liney and Heide 2019). These techniques have been developed to improve treatments by maximizing the dose to the tumour while also sparing surrounding tissues. However, they also increase the complexity of dose measurements and highlight the limitations of existing dosimeters (Ezzell *et al* 2003, Benedict *et al* 2010, Solberg *et al* 2012, Jelen and Begg 2019, O'Brien *et al* 2018).

Considering the dosimetry challenges of recent radiotherapy treatments, plastic scintillation detectors are considered well-suited tools due to their high spatial resolution, fast response, water-equivalence and relatively low magnetic field dependency (Beaulieu and Beddar 2016, Madden *et al* 2019, Alexander *et al* 2020, Cumalat *et al* 1990, Therriault-Proulx *et al* 2018). While they show numerous advantages, irradiation of plastic fibres with megavoltage beams also implies production of Cerenkov light (Beddar *et al* 1992, Boer *et al* 1993). This inherent light emission has long been considered a contamination signal due to its multiple dependencies leading to intensity variations that are not directly linked to the dose deposition (Law *et al* 2006, 2007). Consequently, many techniques have been developed over the past years to overcome its influence in the output signal (Frelin *et al* 2005, Archambault *et al* 2006, Lambert *et al* 2008, Liu *et al* 2011).

On the other hand, interest for Cerenkov emission dosimetry has increased over the past decades. While most emerging techniques are focusing on in-vivo Cerenkov imaging (Tendler *et al* 2020, Hachadorian *et al* 2020) or in-water detection with an out-of-field detector (Glaser *et al* 2013, 2014, Pogue *et al* 2015, Meng *et al* 2019, Zlateva *et al* 2019a, 2019b, Yogo *et al* 2020), few are based on Cerenkov light produced in optical fibres (Jang *et al* 2013, Yoo *et al* 2013). Challenges of using Cerenkov signal from an optical fibre arise from the delimitation of a sensitive volume and the signal discrimination of the Cerenkov also produced in the transport fibre itself. The number of Cerenkov photons emitted is directly dependent of the velocity of the electrons travelling through the medium, and hence their energy (Jelley 1958). Thus, there is a proportionality between the intensity of the emitted light and the dose deposited in the irradiated portion of the optical fibre. It is however necessary to determine the intensity of the signal produced in a finite sensitive volume to link it to the deposited dose. Despite this proportionality, the Cerenkov light intensity collected also possesses an angular dependency (Law *et al* 2006, 2007). In fact, only a small fraction of the Cerenkov light produced in the fibre core is captured and transmitted along the transport fiber. The total intensity reaching the photodetector depends on the angle between the fiber axis and the particle path. This results from the optical photons that are emitted in a conical shape with its axis centred along the path of the charged particle. Thus, its angular dependency could be useful to extract additional information that is out of reach when using other dosimeters and can only be obtained with Monte Carlo simulations.

In this study, we introduce a novel hybrid Cerenkov-scintillation detector that exploits Cerenkov light generated in a clear PMMA optical fibre. The detector is designed to simultaneously performed dose measurements and identify the primary photon or electron beam incident angles based on Cerenkov angular dependency. Accordingly, the main objective of the present study is to investigate the angular dependency of a novel hybrid Cerenkov-scintillation detector that exploits Cerenkov light generated in a clear PMMA optical fibre. Moreover, it aims at validating the ability of the detector to perform simultaneous dose and irradiation angle measurements by using Cerenkov signal variations attributable to its angular dependency.

## 2. Materials and methods

### 2.1 Cerenkov radiation

Cerenkov radiation is the emission of optical photons that occurs when a charged particle travels through a dielectric medium with a velocity greater than the local velocity of light (Cherenkov 1934, Jelley 1958). The resulting optical photons are emitted in the shape of a cone with the path of the charged particle as its axis and having a half-opening angle $\theta$ defined by

$$\cos\theta = \frac{1}{\beta n}, \qquad (1)$$

where $n$ is the refractive index of the medium and $\beta$ is the relativistic phase velocity of the particle. Considering an optical fibre having a PMMA-based core with refractive index of 1.49 (at 520 nm), the minimal energy required for an electron to generate Cerenkov radiation is 178 keV. Thus, external beams from medical linear accelerator (LINAC) composed of photons or electrons will induce in the exposed optical fibre and surrounding media a polyenergetic electron fluence spectrum of which most of the electrons possess an energy greater than the threshold (McLaughlin *et al* 2018, Konefał *et al* 2015). The Cerenkov yield, that is the number of optical photons $dN$ in the wavelength interval between $\lambda_1$ and $\lambda_2$ emitted per unit path length $dl$ of a charged particle having an energy greater than the threshold, is given by the Frank–Tamm formula (Tamm and Frank 1937)

$$\frac{dN}{dl} = 2\pi\alpha z^2 \left(1 - \frac{1}{\beta^2 n^2}\right)\left(\frac{1}{\lambda_2 - \lambda_1}\right), \qquad (2)$$

where α is the dimensionless fine-structure constant and $z$ the charge of the particle. According to Equation (2), the number of optical photons emitted within the fibre increases with the particle energy and the refractive index. The same also applies to the Cerenkov cone half-opening angle $\theta$. However, influence of the refractive index variation of PMMA over the visible spectrum (i.e., 400-700 nm) is considered negligible (Zhang *et al* 2020). Therefore, Cerenkov light emission intensity and angle mostly depend on the energy fluence spectrum. As the dose deposited in the medium by the electrons also depends on their energy spectrum, normalizing the Cerenkov signal to the measured dose should allow to eliminate this dependency. Consequently, signal variations attributable to angular dependency could be used to extract the irradiation angle with an angular calibration curve performed under fixed dose irradiations.

*2.2 Detector conception*

The hybrid Cerenkov-scintillation detector designed for this study was composed of two distinct probes as shown in Figure 1. A first fibre-based Cerenkov detector was built similarly as a single-point plastic scintillation dosimeter (PSD). A 10 mm long optical fibre with a PMMA-based core and fluorinated polymer-based cladding (1 mm diameter ESKA GH-4001, Mitsubitshi Chemical Co., Tokyo, Japan) was used as a finite sensitive volume. The latter was separated by an absorptive filter from a 17 m long transport fibre (ESKA GH-4001) to modify the Cerenkov spectrum produced in the sensitive volume, thus allowing for an algorithm to properly eliminate the contamination signal from the transport fibre. An absorptive filter was preferred to guarantee no light emission was reflected towards its emission source. To increase the total signal collected, the same type of fibre was used to generate and conduct the Cerenkov light to a photodetector due to its high core-cladding refractive index difference $\Delta n$ of 0.11. All components were bonded together using an optical adhesive and sealed with a 3D printed (SL1 printer, Prusa, Prague, Czech Republic) light tight tube made of photo-hardening polyresin. An independent similarly built PSD composed of a 10 mm long BCF12 scintillating fibre (Saint-Gobain, Hiram, USA) was also used simultaneously to provide a real-time dose measurement and account for the Cerenkov electron energy dependency.

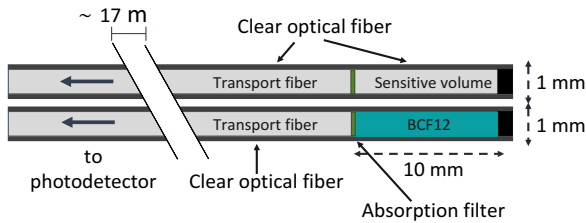

*Figure 1 : Design schematic of the hybrid detector showing the Cerenkov and the scintillation probe assemblies.*

A second hybrid Cerenkov-scintillation detector (Detector 2) was also built to evaluate the influence of the acceptance cone on the Cerenkov signal collected and angular dependency. The only difference between the two detectors lies in the sensitive volume optical fibre (1 mm diameter ESKA MH4001, Mitsubitshi Chemical Co., Tokyo, Japan) having a numerical aperture of 0.3 instead of 0.5. While the latter was used thorough the entire characterization, the former only serve for comparison measurements. Subsequent references to the detector refer to the first prototype with a numerical aperture of 0.5 unless otherwise stated.

*2.3 Signal acquisition set-up*

Irradiations of the detector were carried out using a Varian Clinac IX (Varian Medical Systems, Palo Alto, USA) linear accelerator. For all measurements, 6 and 18 MV photon and 6, 9, 12, 16 and 20 MeV electron beams were used. Absolute doses and dose-rates were validated using a Farmer ionization chamber (TN 31013, PTW, Freiburg, Germany). The multi-channel HYPERSCINT scintillation dosimetry platform (HYPERSCINT RP-200, Medscint Inc., Quebec, Canada) was used to collect simultaneously the light signal emitted by both probes of the detector (Jean *et al* 2021). The transport fibres of 17 m long allowed to place the HYPERSCINT platform outside the treatment room to minimize noise. All measurements were achieved using a wavelength range set from 350 to 635 nm which represents an effective area of 2260 pixels wide on the photodetector. To reduce the readout noise, a binning was performed in the vertical direction of the sensor on 100 pixels for each channel and the cooling system was set at -5° C to keep the temperature stable across all measurements. Each acquisition was made using a repetitive 2 s integration time. Background exposures with matching exposure time were also taken and subtracted from the acquired signal.

*2.4 Signal unmixing and dose calibration*

The signal of both probes was collected and unmixed independently using the hyperspectral approach (Archambault *et al* 2012). The measured spectrum ($m$) is assumed to be a linear superposition of the normalized spectra ($r_i$) of the different light emitting sources and pixels of the detector array are considered as $L$ individual measurement channels to which are assigned wavelengths ($\lambda_j$). This can be expressed in matrix form such as

$$\begin{bmatrix} m_{\lambda 1} \\ m_{\lambda 2} \\ \vdots \\ m_{\lambda L} \end{bmatrix} = \begin{bmatrix} r_{1,\lambda 1} & r_{2,\lambda 1} & \cdots & r_{n,\lambda 1} \\ r_{1,\lambda 2} & r_{2,\lambda 2} & \cdots & r_{n,\lambda 2} \\ \vdots & \vdots & \vdots & \vdots \\ r_{1,\lambda L} & r_{2,\lambda L} & \cdots & r_{n,\lambda L} \end{bmatrix} \begin{bmatrix} x_1 \\ x_2 \\ \vdots \\ x_n \end{bmatrix}. \qquad (3)$$

where $x_i$ represents the contribution of each light emission sources $i$. As the BCF12 and the clear fibre probes are constructed as single-point PSD, only the scintillation and the filtered Cerenkov from the sensitive volume contributes to the measurement, respectively. The total measured spectrum also includes two contamination signals from the transport fibre that is exposed to the beam, which are unfiltered Cerenkov and fluorescence (Nowotny 2007, Boer *et al* 1993, Therriault-Proulx *et al* 2013). Thus, the variable *n* in this study is equal to 3 to account for all light emission sources.

To solve the system for the variable *x*, the left pseudo-inverse matrix method is used as follows

$$x = (R^T R)^{-1} R^T m. \quad (4)$$

The raw spectrum of each element that contributes to the total signal measured is required to solve Equation (4). The fluorescence spectrum was obtained by placing 60 cm of rolled transport fibre directly on the exit window of the On-Board Imaging kV source arm (OBI, Varian Medical Systems, Palo Alto, USA). The latter was set at 120 kVp for a 40 second irradiation. As for the raw spectrum of the Cerenkov probe sensitive volume, it is not possible to use a beam with lower energy than the Cerenkov threshold, contrary to plastic scintillator detectors. Accordingly, a procedure similar to the one described by Guillot et al. (Guillot *et al* 2011) was used. As illustrated in Figure 2, the probe was irradiated with a LINAC in 3 respective conditions. C2 and C3 measurements are performed using an irradiation angle where the fluorescence to Cerenkov ratio is minimized (i.e., at the maximal Cerenkov emission angle) while the C1 measurement is performed using a known dose at normal incidence with respect to the fibre axis. The difference between calibration measurements C3 and C2 is only due to contamination signal as the dose received to the sensitive volume in both conditions is identical. Therefore, the contamination signal is then used to unmix the C1 measurement signal and hence extract the sensitive volume raw spectrum.

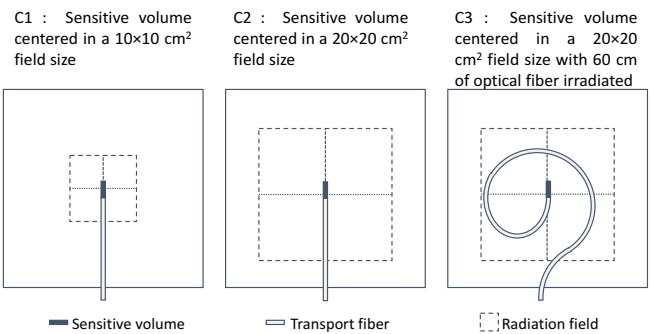

*Figure 2 : Schematic top view of the three different configurations used for the calibration measurements with a LINAC.*

The intensity $(x_{i,C1})$ calculated for the deposited dose $(d_{i,C1})$ with irradiation C1 is then used to determine the dose received in other irradiation conditions $(d_i)$ such as

$$d_i = d_{i,C1} \frac{x_i}{x_{i,C1}}. \quad (5)$$

While the equation (5) is valid for any irradiation conditions when using a scintillation detector, nominal energy, angle and depth of measurement must match those of the calibration irradiation for a Cerenkov detector used for dosimetry purpose. Consequently, the Cerenkov detector signal was dose calibrated at a $d_{max}$ depth (i.e. 1.5 cm at 6 MV, 3.5 cm at 18 MV, 1.5 cm at 6 MeV, 2 cm at 9 MeV and 3 cm at 12, 16 and 20 MeV) and at normal incidence for this study.

### 2.5 Dose linearity, output factors and dose rate independence

First measurements aimed at evaluating the efficiency of the absorptive filter to produce a sufficiently distinct Cerenkov spectrum allowing for the algorithm to properly eliminate the signal from the transport fibre. Given that, the Cerenkov probe was used in the same manner as a PSD. Both probes were embedded in a solid water phantom at a $d_{max}$ depth with their sensitive volume at the isocentre, and a 10 cm thick slab was placed underneath to provide backscatter. Irradiations were performed at normal incidence with respect to the fibre axis (i.e., gantry at 0˚ with the probes along the lateral axis)

Dose linearity measurements were achieved using the clear fibre probe signal as a function of the deposited dose measured by the scintillator. Irradiations consisted of various doses ranging from 20 cGy to 500 cGy in a 10x10 cm$^2$ field size. Then, output factors were measured for field sizes varying from 5×5 to 25×25 cm$^2$ using 3 repeated irradiations of 100 cGy for each energy. To calculate the relative outputs, the signal measured with both probes at field size $n×n$ cm$^2$, using jaw-defines fields for photons and applicator size for electrons, was normalized to the signal measured under a 10×10 cm$^2$ field.

Validation of the dose rate independence was achieved using a 10x10 cm$^2$ field size with the surface-to-source distance (SSD) gradually increased to reduce the mean dose rate. For each couch position, respectively five irradiations of 200 cGy at 600 MU/min were realised under 6 and 18 MV photon beams only. To account for the field size variation, attenuation in air, and treatment room backscattering at high SDD, absolute dose rate measurements relative to SSD were performed with a TN 31013 ionization chamber, to which the Cerenkov probe measured doses were then normalized.

### 2.6 Angular dependency

A validation of the angular dependency was obtained with irradiations ranging from 20 cGy to 500 cGy using a 10×10 cm$^2$ field size. Measurements were repeated at various gantry angles between 0˚ to 90˚. All measurements were carried out with the two probes inserted in solid cylinder-shaped water-equivalent phantom to provide a scattering medium and ensure a constant depth of measurement regardless of the gantry angle. The phantom was manufactured using a 3D-printer (MK3S+ and SL1, Prusa, Prague, Czech Republic) and was composed of a methacrylate polymers resin (Polyresin-Tough, Prusa Polymers, Prague, Czech Republic). A first phantom of 6 cm diameter was used for the photon beams as well as the 12 to 20 MeV electron beams. A smaller cylinder of 3 cm diameter was used for lower electron energies to provide a depth of measurement closer to the $d_{max}$ depth. As illustrated in Figure 3, the phantoms were suspended in the air on a stand at the foremost longitudinal position to avoid any beam attenuation from the treatment table. Both probes were placed along the lateral axis with their sensitive volumes pointing toward the X2 jaw.

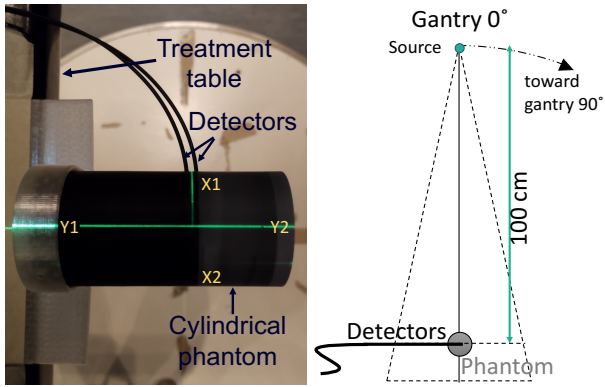

*Figure 3 : (Left) Top view of the set-up used for the validation of the Cerenkov prototype showing the 2 probes inserted in the 3D printed cylindrical phantom placed on its stand. (Right) Schematic of the phantom showing the orientation of the probes with respect to the gantry position.*

## 2.7 Angular calibration curve

Calibration curves of the Cerenkov signal intensity as a function of the radiation incident angle for fixed doses of 200 cGy were performed in a 10×10 cm² field size. For the photon beams, the whole 360 degrees was covered using increments of 5 degrees. For the electron beams, the applicator hinders the gantry from reaching angles ranging between 215° and 325° without damaging the transport fibre. Thus, all other angles were covered using increments of 5 degrees with electron beams. With the detector placed at the isocentre, the irradiation angle corresponds to the gantry angle. Each calibration curve was then normalized to the deposited dose such as

$$I_{\theta,Calib} = \frac{x_{\theta,Calib} \times d_{C1}}{200\ cGy \times x_{C1}}, \qquad (6)$$

where $x_{\theta,Calib}$ is the Cerenkov probe signal at a given angle and $d_{C1}$ and $x_{C1}$ are respectively the dose delivered and Cerenkov intensity both measured in dose calibration conditions (i.e., at a depth of d_max in a 10×10 cm² field size for a normal incidence). Doses $d_{C1}$ delivered in calibration conditions were validated using the ionization chamber while the 200 cGy was measured using the BCF12 scintillator signal.

## 2.8 Irradiation angle measurement set-up for uneven dose

The ability of the prototype to correctly measure the irradiation angle using both signals was validated by varying simultaneously the dose and irradiation angle. The detector was placed 20 cm below the isocentre in a field size of 40×10 cm² for the photon beams. As the SSD and the off-axis distance vary according to the gantry position, all irradiation angles provided different mean dose rates, hence a non-constant dose. The detector being shifted from the isocentre also provided an irradiation angle that differs from the gantry angle. This ensured that conditions for both dose and irradiation angle were different from the calibration curve measurements. As for the electrons, the 20×20 cm² applicator prevented the phantom to be at an SSD smaller than 97 cm for any angles between 0° and 180°. Accordingly, the detector was placed at the isocentre height. The phantom was instead shifted 8.5 cm toward the X1 jaw direction on the lateral axis as shown in Figure 4. Irradiations of 200 MU for gantry angles ranging between 0° and 180° by increments of 5° were performed for all energies and modalities.

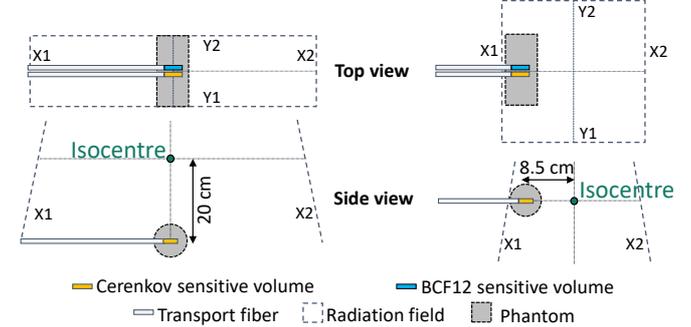

*Figure 4 : Schematic top and side views of the set-up used for irradiation angle measurements using photon and electron beams showing the coordinate system. The source-detector distance (SDD) is calculated with the gantry positioned at 0°.*

## 2.9 Irradiation angle calculation

The irradiation angle is calculated using a 3-steps process. First step consists of measuring the deposited dose in the Cerenkov sensitive volume at each gantry angle using the scintillation probe signal. Then, the Cerenkov probe signal ($x_\theta$) is normalized to the dose measured by the BCF12 scintillator ($d_{\theta,scint}$) delivered at an irradiation angle $\theta$. This normalization also includes a Cerenkov dose-light calibration factor such as

$$I_\theta = \frac{x_\theta \times d_{C1}}{d_{\theta,scint} \times x_{C1}}, \quad (7)$$

where $d_{C1}$ and $x_{C1}$ are respectively the dose delivered and Cerenkov intensity obtained by dose calibration measurements. A new dose-light calibration factor is required at each new measurement session to account for any signal losses resulting from the fibre bending, connectors and mechanical and radiation induced damages that could also affect the collected signal. Finally, the Cerenkov angular calibration curve is interpolated using a spline polynomial fit to obtain a calibration function such as

$$f(\theta) = I_{\theta,Calib}. \quad (8)$$

The irradiation angle is obtained by solving the calibration function for the variable $\theta$ using the Cerenkov intensity values normalized to the dose ($I_\theta$). Due to the symmetry of the Cerenkov angular dependency, one intensity value can have up to four different angles as solutions. Consequently, conditions were set to obtain a single angle value per given intensity. It was assumed that the starting angle and the direction of rotation were both known, and that the latter did not change during an acquisition. Therefore, the first irradiation angle was solved by searching a solution nearby the starting angle. Subsequent angles are solved with the condition of being greater or lesser than the preceding measured angle for clockwise or counterclockwise rotation, respectively.

As the gantry angles were not equal to the incident angles due to the phantom being shifted from the isocentre, the latter were then calculated for each acquisition using the measurement conditions and set-up geometry. The source-isocentre and detector-isocentre distances being known, the irradiation angle was determined using laws of sines and cosines. For this calculation, the radiation produced by the target was assumed to be a point source. While the Cerenkov signal intensity is directly linked to the angular distribution of all the electrons above the Cerenkov production threshold energy, the mean incident angle of the primary beam was used to produce the calibration curves. Accordingly, calculated irradiation angles allowed to validate the accuracy of the measured irradiation angles using the hybrid Cerenkov-scintillation detector.

## 2.10 Influence of the numerical aperture on the Cerenkov signal

The effect of the sensitive volume numerical aperture (NA) on the Cerenkov signal collected was investigated using the Detector 2 having a smaller critical angle of 17°. The latter was used to perform intensity measurements of the Cerenkov signal at normal and parallel incidence but also at the peak intensity angle by delivering 200 cGy. Three repeated measurements were carried out for each irradiation conditions.

For all modalities, signals measured at normal and parallel incidences with respect to the fibre axis were normalized to their respective peak intensities. All ratios were then compared to those of the Detector 1, which required no further measurements as they were extracted from the angular calibration curves obtained previously. A complete calibration curve and irradiation angle were also measured with the Detector 2 at 6 MeV in the same manner as described in the preceding sections.

## 3. Results

### 3.1 Signal unmixing

The three spectra composing the signal measured by the Cerenkov probe, including the filtered Cerenkov from the sensitive volume as well as the two contamination signals, are shown in Figure 5. The signature shape of the sensitive volume spectrum results from the notch absorptive filter that was chosen to build the prototype.

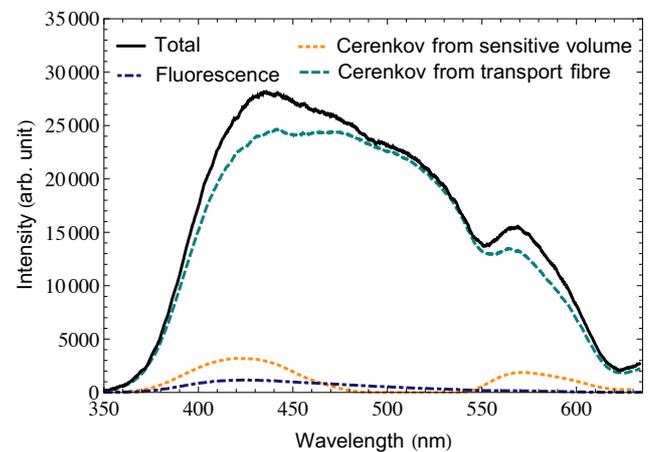

*Figure 5 : Total intensity measured for a 100 cGy irradiation and individual spectra of each light emitting element of the Cerenkov detector. The shape of the filtered Cerenkov spectrum results from the absorptive filter. All spectra are affected by the wavelength response of the photodetection system and the transport fibre transmission spectrum.*

An example of the Cerenkov probe total intensity measured under a 100 cGy irradiation at normal incidence and the different light emitting source contributions are illustrated in Figure 5. A portion of the total measured spectrum includes some intensity due to fluorescence produced in the transport fibre as the fluorescence to Cerenkov ratio was greater than the calibration measurement conditions. The contribution of the 1 cm long sensitive volume represents only a slight fraction of the total intensity collected as the probe was irradiated with a field size of 10 x 10 cm$^2$.

### 3.2 Dose linearity, output factors and dose-rate independence

The signal intensity of the Cerenkov sensitive volume was found to follow a linear trend as a function of the dose measured with the scintillation signal for both modalities used. Results obtained for the signal characterization of the Cerenkov probe under 6 and 18 MV photon beams at normal incidence for various doses are shown in Figure S1.a in the supplementary material section (S.M.). Results for the same measurements using 6 to 20 MeV electron beams are presented in Figure S1.b. The influence of the energy on the measured intensity was found to vary according to the radiation beam type. While the intensity collected for a given dose tends to increase with energy for the photon beams, an opposite tendency was observed for the electrons. As the Frank-Tamm formula predicts a greater yield with increasing energy, this observation is only valid at normal incidence and the trend at the peak intensity angle should obey the predicted energy dependency.

Compared to ionization chamber measurements, relative output factors were accurately measured within ±0.8 % using the Cerenkov probe at 6 and 18 MV. The primary particle type did not affect the dose measurements as the Cerenkov detector provided similar accuracy with various electron energies. The average accuracy of the Cerenkov detector is similar to what was observed with the BCF12 detector. Table S1 of the S.M. section displays relative output factors obtained with the Cerenkov and the scintillation probes under 6 and 18 MV photon beams while electron beam results are presented in Table S2.

At normal incidence, the measured dose using the Cerenkov sensitive volume signal was found to be dose-rate independent, as expected. For fixed dose irradiations, the sensitive volume signal shows a discrepancy with the predicted dose that reaches 1.1 % at 6 MV for the lowest dose rate tested. Similar results were obtained under an 18 MV photon beam and the dose was accurately measured within ±0.7 %. Dose ratio of the Cerenkov detector to the ionization chamber as a function of the mean dose rate can be found in Figure S2 of the S.M. section.

*3.3 Angular dependency*

The signal intensity of the Cerenkov probe as a function of the dose measured with the scintillation probe at various angles using two photon beam energies are presented in Figure S3 of the SM section. While the linear dose-light relationship can be observed for the whole range of angles tested at both energies, the slope of the regression varies based on the irradiation angle due to the Cerenkov angular dependency. Same results were also obtained with the electron beams (not shown). For all gantry angles and energies tested, the lowest regression slope coefficient of determination was found to be 0.99991.

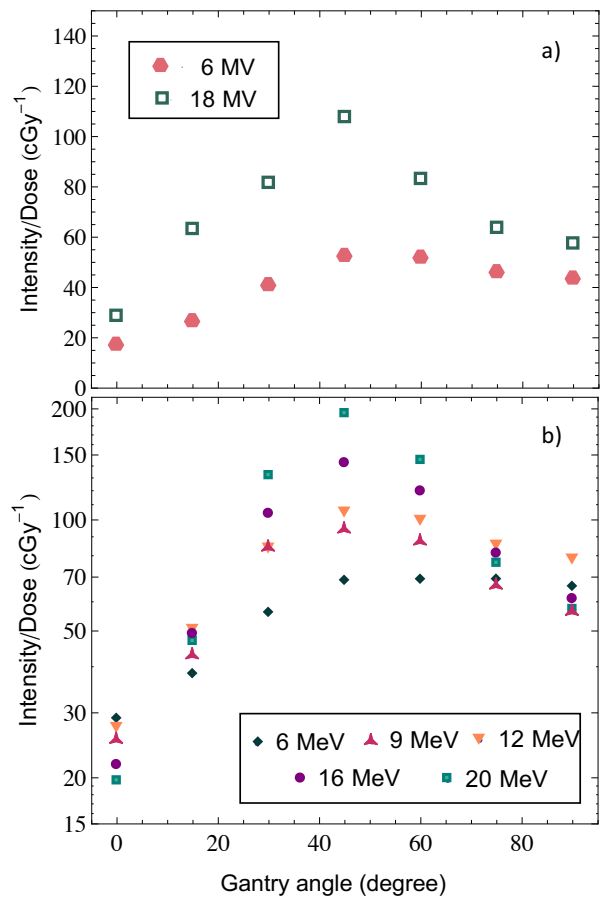

*Figure 6 : Ratios of the signal emitted by the Cerenkov detector sensitive volume to the deposited dose as a function of the gantry angle obtained with (a) 6 and 18 MV photon and (b) 6, 9, 12, 16 and 20 MeV electron beams. Error bars are smaller than the data point symbols.*

The signal intensities of the Cerenkov probe per unit of dose measured with the scintillation probe as a function of the irradiation angle are presented in Figure 6.a and 6.b for photon and electron beams, respectively. Comparing the results at various energies shows that the angular dependency also varies as a function of the beam energy. For the photon beams, the intensity per dose unit increases as a function of the energy for all angles. For the electron beams, the intensity collected for a given dose at a gantry angle of 0° tends to decrease with energy as illustrated in Figure 6.b. However, the trend at the peak intensity angle obeys the predicted dependency with a greater yield for increasing energy.

*3.4 Angular calibration curve*

Signal emitted by the Cerenkov sensitive volume and the scintillator as a function of the irradiation angle for fixed dose are illustrated in Figure 7.a, 7.b and 7.c for the various photon and electron beam energies. All signals were normalized to their maxima to emphasize the influence of the beam energy on the angular dependency. For both modalities, the increased beam energy was found to narrow the captured Cerenkov emission cone. Consequently, the 18 MV, 16 and 20 MeV angular calibration curves showed greater slopes at normal incidence and at the peak intensity angle. However, the 9 MeV curve differs from the others as its slope is greater than at 12 MeV. This can be linked to the phantom that is smaller for the 9 MeV beam, thus reducing the scattered electron contribution. Regarding the scintillation (shown by the dashed lines on Figure 7), the intensity as a function of the angle was found to be isotropic for all modalities due to the absence of angular dependency, as expected.

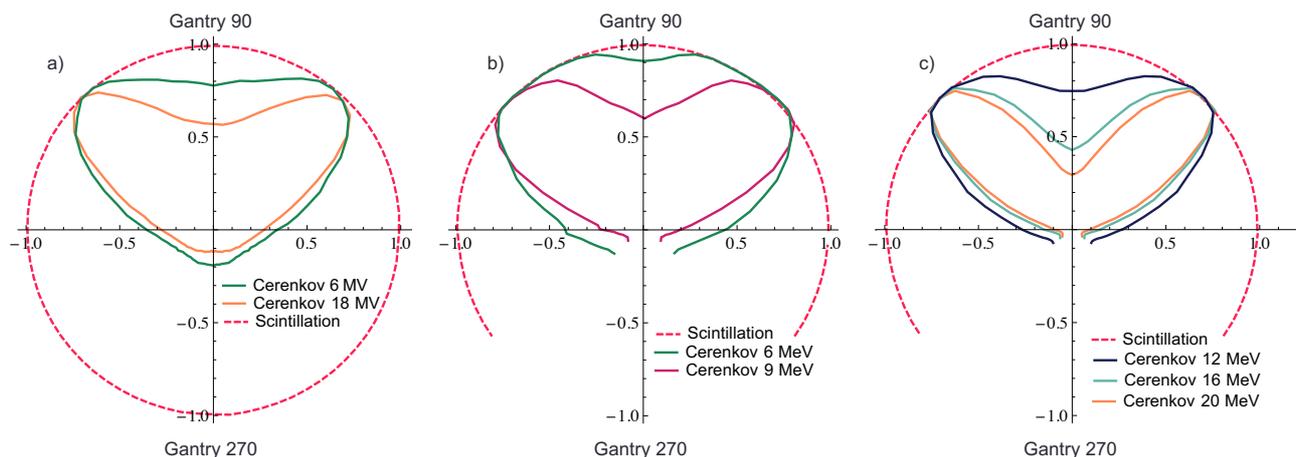

*Figure 7: Signal emitted by the Cerenkov detector sensitive volume and the BCF-12 scintillator as a function of the angle of incidence for fixed dose using (a) 6 and 18 MV photon, (b) 6 and 9 MeV and (c) 12, 16 and 20 MeV electron beams. Figure (a) and (c) were obtained using the 6 cm diameter phantom while the 3 cm diameter phantom was used for (b). All signals are normalized to their respective maxima*

### 3.5 Cerenkov intensity for uneven doses

Examples of the Cerenkov signal measured as a function of the gantry angle for uneven doses using 6 MV and 20 MeV beams are presented in Figure 8.a and 8.b, respectively. The dose that was measured simultaneously by the scintillator is shown in a blue dashed line. The difference between the two detector signals results from the Cerenkov angular dependency which leads to intensity variations that are not directly linked to the dose deposition. As the acquisition set-up was different with the electron beams, the measured dose as a function of the irradiation angle differed from the photon beams. Unfortunately, the dose variation was also limited by the set-up due to the electron applicator.

### 3.6 Irradiation angle measurements

Figure 9 displays an example of the Cerenkov probe intensity normalized to the dose obtained with the scintillation probe at each gantry position using a 6 MV beam. The Cerenkov intensity without the dose normalization is also included to show the dose dependency effect on the signal. As the gantry angle does not correspond to the irradiation angle due to the phantom positioning, a shift from the angular calibration curve can be observed, as expected.

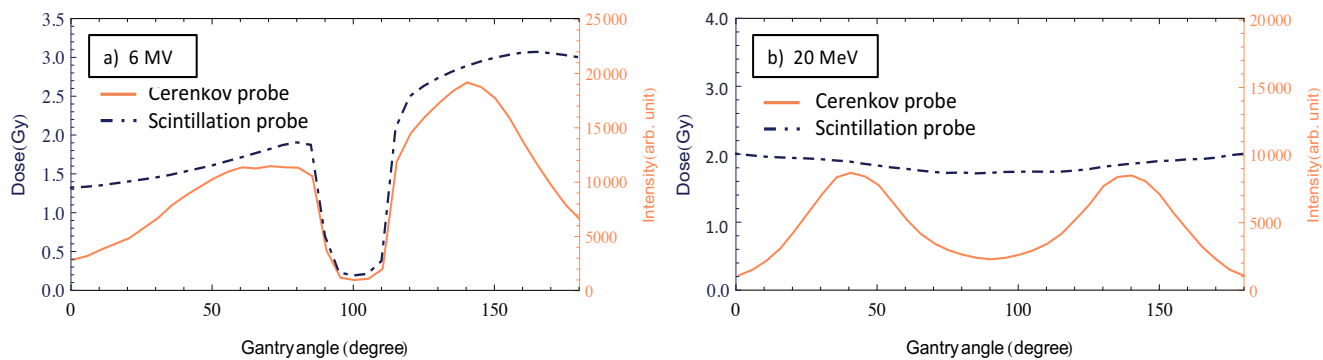

*Figure 8: Interpolated dose measured with the scintillation probe previously dose calibrated and the signal measured by the Cerenkov probe as a function of the gantry angle using (a) 6 MV photon and (b) 20 MeV electron beams. The Cerenkov signal variations are due to both dose and irradiation angle dependencies.*

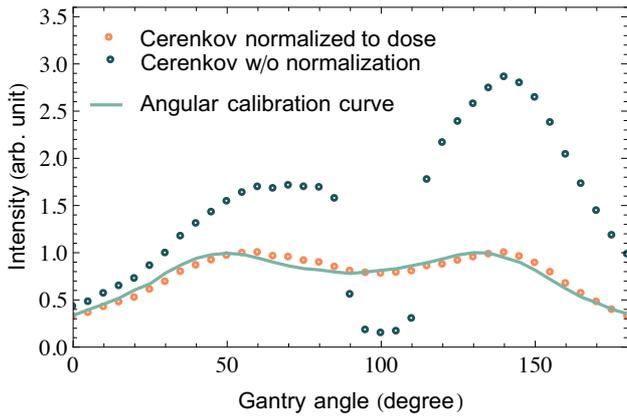

Figure 9: Cerenkov intensity measured as a function of the gantry angle with and without normalization to the dose measured by the scintillator at 6 MV. The calibration curve is also included for comparison. Each point corresponds to an individual measurement.

The irradiation angle absolute errors obtained by solving the calibration function with the Cerenkov signal normalized to the dose are presented in Figure 10 for the photon and electron beams. The difference was obtained by using the real incident angle calculated with the set-up geometry and the source position. Irradiation angle measurements exhibit an absolute mean error of 1.86° and 1.02° at 6 and 18 MV, respectively, as shown in Figure 10.a and 10.b. Similar results were obtained with the four different electron beam energies as depicted in Figure 10.c, 10.d, 10.e and 10.f. The absolute mean error reaches 1.97°, 1.66°, 1.45° and 0.95° at 9, 12, 16 and 20 MeV, respectively. For the photon beams, the lowest doses are linked to the highest differences observed while greatest discrepancies are located around the peak intensity angle for the electron beams. A repetitive structure in the absolute error plots resulting from the four subfunctions defining the piecewise polynomial fit can be observed. For all measurements, error bars were calculated using the mean error on the Cerenkov signal intensity.

### 3.7 Numerical aperture influence

Signal of the two different Cerenkov probes at normal and parallel incidences with respect to the fibre axis normalized to their respective peak intensities are depicted in Figure 11. While the decreased numerical aperture was found to have little influence on the collected intensity ratios at high energies, a pronounced effect at lower energies can be observed. For both modalities, ratios at normal and parallel incidences at the lowest energies were significantly reduced with the probe having a NA of 0.3 compared to the 0.5.

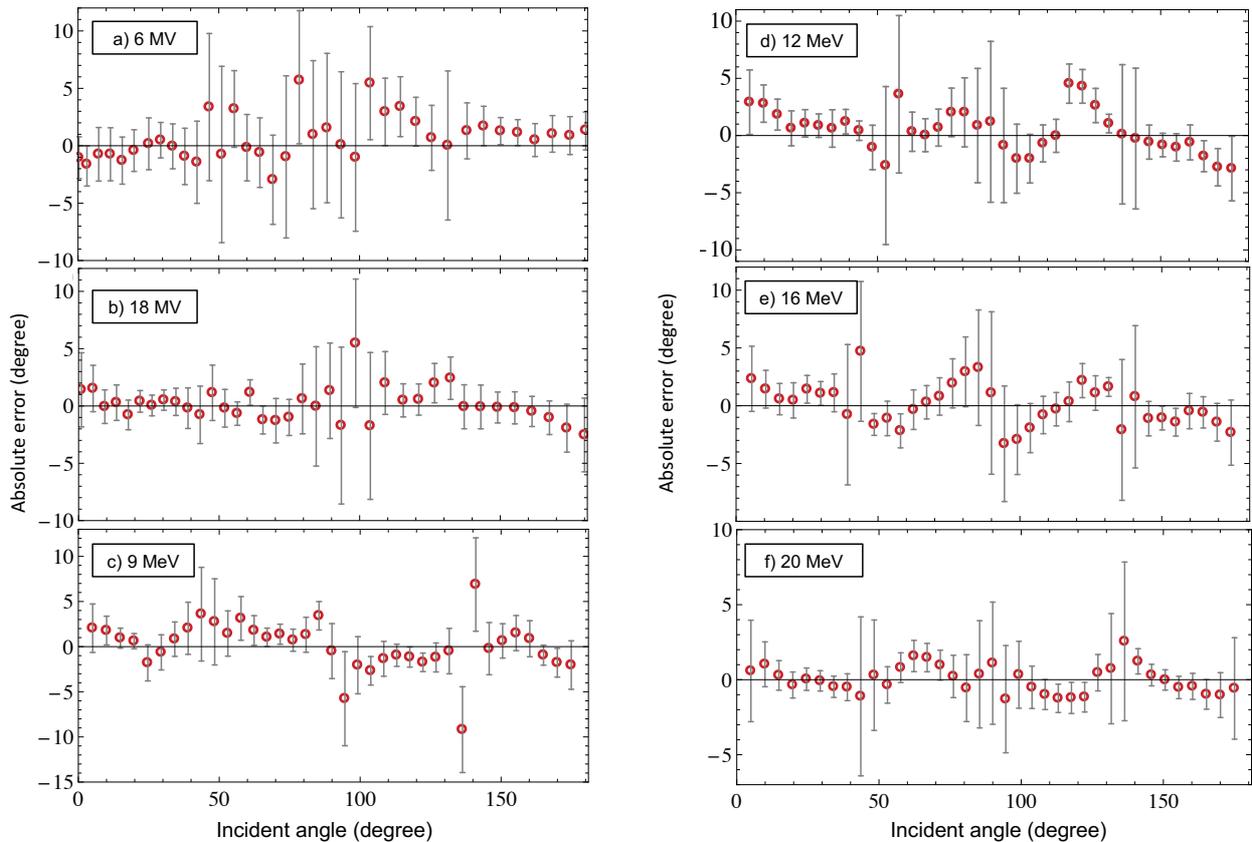

Figure 10 : Angle of incidence absolute error measured with the Cerenkov detector using a dose varying as a function of the irradiation angle for a) 6 MV and b) 18 MV photon beams and (c) 9 MeV, (d) 12 MeV, (e) 16 MeV and (f) 20 MeV electron beams.

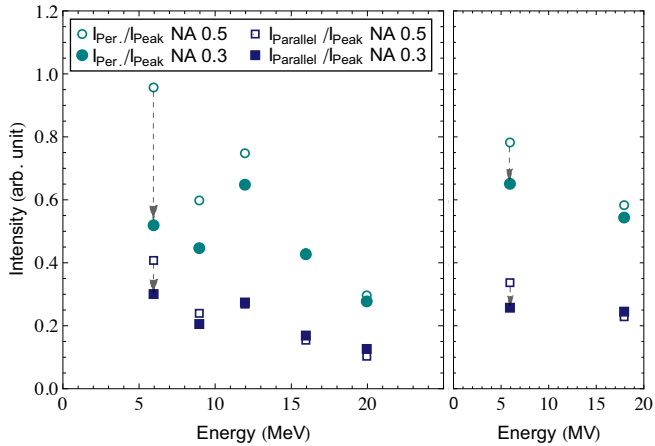

*Figure 11: Ratios of the Cerenkov sensitive volume signal measured at parallel and normal incidence to the peak intensity as a function of the energy using the two detectors with numerical aperture of 0.3 and 0.5. Error bars are smaller than the data point symbols.*

Angular calibration curves of a 6 MeV beam with both Cerenkov probes have demonstrated that reducing the NA emphasizes the angular dependency (see Figure S4 of the SM section). Narrowing of the fibre acceptance cone has lessened the portion of the captured Cerenkov emission cone, therefore enhancing the slope of the curve. Thus, irradiation angle measurements at 6 MeV are now possible with Detector 2. It yields a mean absolute error of 4.82°, with the distribution as function of angle of incidence illustrated in Figure 12. Discrepancies can be seen around the peak intensity values similarly to the other electron energies. However, the Cerenkov detector using a 6 MeV beam misrepresents all angles measured below 40° and above 140°.

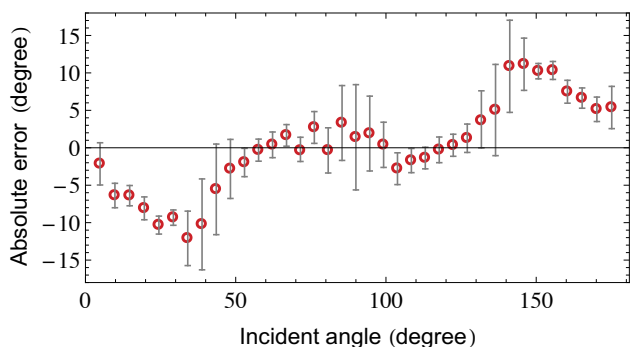

*Figure 12: Angle of incidence absolute error measured with the Cerenkov detector (NA=0.3) using a dose varying as a function of the irradiation angle for 6 MeV*

## 4. Discussion

### 4.1 Cerenkov probe dose-light relationship

Characterization of the relationship between the Cerenkov probe sensitive volume signal collected and the deposited dose is of great interest to validate the suitability of the detector. As the Cerenkov probe was used in the same manner as a PSD at first, it allowed to validate the proportionality of the Cerenkov signal and the dose for all modalities and energies tested. While there is no direct dependency between the two of them, they both share an electron energy spectrum dependency. Due to this dependency of the Cerenkov yield, the capture and transmission along the fibre are affected by the depth of measurement. As electron energy loss increases with the travelled distance, the slowing-down process induces fewer Cerenkov photons with lower angle between the fibre axis and the particle path due to the narrowing of the cone opening. However, results showed that using calibration conditions where the electron energy spectrum is identical to the measurement conditions allows to rely on deposited dose to account for this dependency.

### 4.2 Contamination signal influence

We investigated the accuracy of output factor measurements using the Cerenkov probe and did a comparison with the scintillator for a wide range of field sizes. The reason for this test was to compare the efficiency of the Cerenkov signal separation method as the contamination signal increases. While the algorithm worked properly for a constant contamination to sensitive volume signal ratio, it was necessary to validate that an increase of the contamination signal resulting from larger field sizes would not affect the Cerenkov probe accuracy. It was found that both detectors provide similar accuracy for all field sizes and modalities tested. This result is a pre-requisite as the dose measured by the scintillation detector is intended to eliminate the electron energy spectrum dependency of the Cerenkov signal. Furthermore, the dose rate independence measurements were also conducted with increasing field size as a function of the SSD. While the dose was accurately measured within ±0.8% with the Cerenkov probe at 18 MV, a greater difference was observed with the 6 MV beam at the lowest dose rate. This discrepancy arises from the total signal that decreases as the electron fluence is reduced for larger SSD.

### 4.3 Energy and angular dependency

The nominal energy of the beam has shown to influence the total intensity collected as a function of dose. In agreement with the Frank-Tamm formula, the 18 MV beam induced a greater amount of Cerenkov light in the sensitive volume than at 6 MV. At the peak intensity, that is approximately twice the number of optical photons for a given dose that are collected using the 18 MV beam. Moreover, the intensity per Gy as a function of the irradiation angle was also found to vary according to energy. The most significant differences between the 6 and 18 MV beam measurements can be observed for gantry angles ranging between 45° and 90° (Figure 6). While a greater variation of intensity per dose unit as a function of the irradiation angle can be observed at 18 MV for this interval, the regression slope values obtained at 6 MV only showed minor changes. This results from the increase of electrons having lower energy which consequently shifts the angular dependency toward lower angles between the fibre axis and the particle paths. Thus, it allows more of the Cerenkov cone to be captured. The difference between the angular dependency of the two energies further arises from the particle paths. In fact, the secondary electrons resulting from Compton effect at higher energies tends to be scattered forward (Andreo *et al* 2017). Accordingly, the Cerenkov angular dependency at 18 MV will be increased as the particle paths better reflects the irradiation angle.

For the electron beams, measurements performed at normal incidence with respect to the fibre axis (i.e. gantry at 0°) shows that the intensity collected per dose unit declines as a function of the energy. This is counter-intuitive as the Cerenkov yield is supposed to increase with energy according to Frank-Tamm formula. This irregularity at normal incidence is also attributable to the electron paths that tend to go forward at higher energies. Contrarily, lower energy beam provides numerous electrons with arbitrary trajectories that are not related anymore to the beam incident angle. Thus, the Cerenkov angular distribution is broader, and it increases the amount of light reaching the detector at lower angles. Inversely, the signal at higher energies decreases due to the narrowing of the Cerenkov cone. At the peak intensity angle, the 20 MeV beam induced a greater amount of Cerenkov light in the sensitive volume in agreement with the predicted Cerenkov yield. Nonetheless, due to the important variations of the intensity per dose unit as a function of the beam energy for both modalities, a calibration must be performed for each of them to enable the use of the detector for dose or angle measurements.

*4.4 Irradiation angle measurements*

Improved technology and complex treatments in modern radiotherapy require a high level of accuracy and consistency. Geometric uncertainties, such as errors on the field position and the irradiation angle relative to target volumes, affect the dose distribution, causing either underdosing of the target volume or overdosing of organs at risk. Therefore, the uncertainty on the gantry angle should be less than 1 degree according to AAPM TG-142 (Klein *et al* 2009). A detector capable of dose and irradiation angle measurements could be useful for either machine or patient specific quality assurance tests to reduce uncertainties and errors in treatment delivery and equipment performances.

The angular dependency of the Cerenkov probe was proven to be a suited tool for irradiation angle measurements. The technique developed in the present study to account for the Cerenkov signal variations attributable to the electron energy spectrum dependency using the deposited dose was found to provide appreciable results. As plastic scintillation detectors have already proven their ability to perform accurate dosimetric measurements, the hybrid detector can rely on the scintillator for dose measurements necessary for irradiation angle measurements. Although most of the angle errors fall within the tolerance imposed by the Cerenkov intensity mean error, few outliers can be seen at all energies. For the photon beams, they correspond to the lowest doses and dose rates arising from the phantom positioning. Accordingly, the total signal collected was relatively low compared to the angular calibration curve. This resulted in higher discrepancies as the accuracy is closely related to the Cerenkov sensitive volume signal intensity. Also, the dose-light calibration factor obtained before irradiation angle measurements showed that the fibre has suffered a 47 % signal drop since its first use. Combination of the fibre bending, connectors, mechanical damages and radiation induced damages from a total dose reaching over 4 kGy have resulted in signal transmission losses. Therefore, a new prototype and an improved connection and bending reproducibility would increase the signal and the detector accuracy.

As for the electron beams, the poor angular dependency of the 6 MeV beam has prevented this energy to be used with the first prototype. Since intensities collected at many angles are identical, the algorithm could not properly calculate the irradiation angle with the conditions imposed and provided an infinite number of solutions. For other energies, outliers lie around the peak intensity angle. Since the calibration curves were performed for gantry angles with increment of 5 degrees, an interpolation was required to obtain a continuous curve. Thus, the calibration curves at angles where the slope changes drastically, such as the peak intensity angle and normal incidence, could be imprecise and therefore causes some of the outliers. Accordingly, a thorough angular calibration curve and signal optimization of the detector could significantly improve the detector performances and help meet AAPM TG-142 tolerance criteria.

*4.5 Numerical aperture*

The numerical aperture was found to affect the angular dependency of the lowest energy beams by reducing the portion of the Cerenkov emission cone captured for both modalities. As the angular distribution of the Cerenkov decreases at higher energy, the acceptance cone of the optical fibre has less impact on the Cerenkov capture. While the irradiation angles could be measured accurately with the detector placed in the central portion of the beam profile at 6 MeV, an increasing off-axis distance has led to higher discrepancies reaching up to 12°. This can be attributable to an important variation in the electron energy spectrum resulting from an increased contribution of scattered electrons compared to the calibration measurements. The penumbra region being wider for low energy beams, this effect is less perceptible for other electron beams used in this study. Moreover, electrons of approximately 1 MeV or less reduce significantly the Cerenkov emission angle (Law *et al* 2006). Consequently, a smaller numerical aperture will have a greater impact on the captured Cerenkov emission for angles approaching normal incidence, which coincides with the greatest off-axis distance in the set-up used for irradiation angle measurements. Due to the complexity of field geometries in patient treatment plans, this limitation of the detector is more likely to prevent its usage with a 6 MeV beam unless a calibration method can account for the electron energy spectrum variations.

## 5. Conclusion

Using a first prototype of a Cerenkov-scintillation detector, the Cerenkov dose-light relationship for fixed angle measurements was validated for 6 and 18 MV photon and 6 to 20 MeV electron beams. Output factors were accurately measured within ±0.8 % for field size up to 25x25 cm$^2$ for all energies tested with both photons and electrons. The average accuracy is similar to what was observed with the BCF12 detector.

The ability to perform simultaneous dose and irradiation angle measurements of external photon and electron beams was also shown. The irradiation angle measurements ranging between 0° and 180° revealed an absolute mean error of 1.86° and 1.02° at 6 and 18 MV, respectively. The accuracy of the detector was also similar while using four different electron beam energies. The absolute mean error was respectively of 1.97°, 1.66°, 1.45° and 0.95° at 9, 12, 16 and 20 MeV. As for the 6 MeV beam irradiation angle measured with the detector having a smaller numerical aperture, the absolute mean error reaches 4.82°. While also measuring the deposited dose, the hybrid Cerenkov-scintillation detector provides useful additional information that other conventional dosimeters are unfit to measure.

First results demonstrate the effectiveness of the proposed detector and irradiation angle measurement method as a potential tool for future quality assurance applications in external beam radiotherapy, especially for stereotactic radiosurgery (SRS) and magnetic resonance image-guided radiotherapy (MRgRT). Moreover, its applications could be extended to brachytherapy treatments as part of a source tracking device. This avenue will be explored in future work.

## Acknowledgements

This work was supported by the *Natural Sciences and Engineering Research Council of Canada* (NSERC) Discovery grants RGPIN 05038-2019 and the Fellowships Program of the *Ministère de la Santé et des Services Sociaux du Québec* (MSSS). François Therriault-Proulx and Simon Lambert-Girard are Co-founder at Medscint inc., a company developing scintillation dosimetry systems. This work was not financially supported by Medscint. The authors thank Benjamin Côté for his help manufacturing the probes and Frédérik Berthiaume for providing technical support with the Hyperscint dosimetry platform.

**Supplementary material**

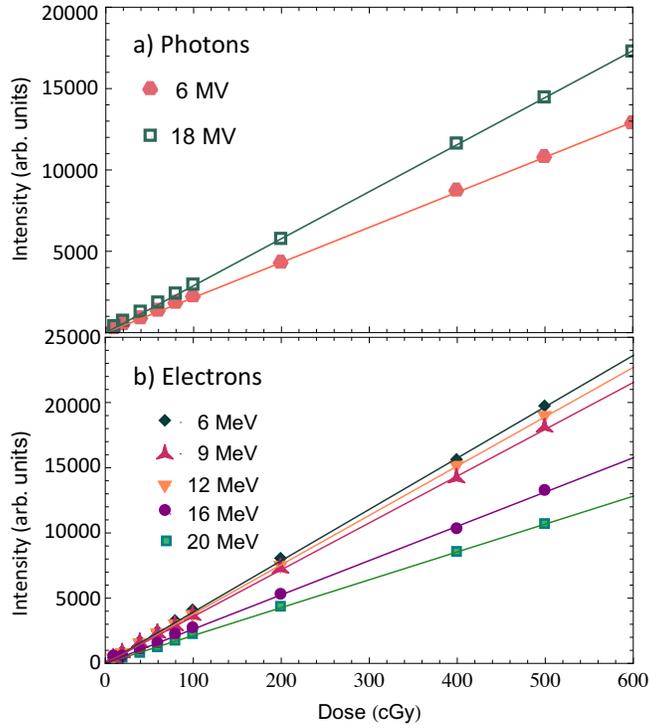

*Figure S1: Signal emitted by the Cerenkov detector sensitive volume as a function of the dose obtained with (a) 6 and 18 MV photon beams and (b) 6, 9, 12, 16 and 20 MeV electron beams at normal incidence. All coefficients of determination are above 0.99991.*

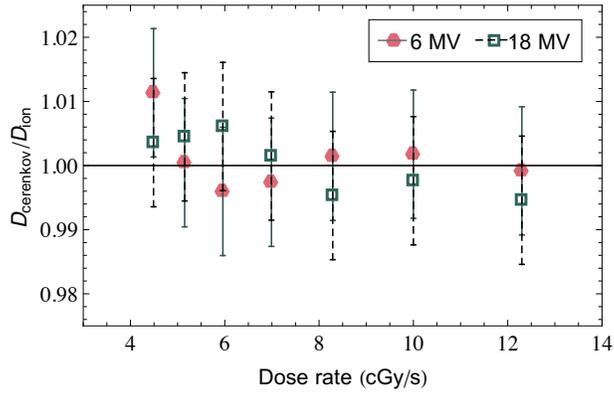

*Figure S2: Measured dose at normal incidence with the Cerenkov detector normalized to an ionization chamber as a function of the dose rate.*

*Table S1: Photon beam relative output factors normalized to a 10 x 10 cm² field size measured with the Cerenkov and scintillation detector in comparison with a TN30013 Farmer ionization chamber (PTW, Freiburg, Germany) and their relative differences.*

| Energy | Field size (cm²) | Ionization chamber | Cerenkov detector | Relative difference (%) | BCF12 detector | Relative difference (%) |
|---|---|---|---|---|---|---|
| 6 MV | 5 x 5 | 0.942 | 0.950 | 0.8 | 0.944 | 0.2 |
| | 10 x10 | 1 | 1 | - | 1 | - |
| | 15 x15 | 1.032 | 1.031 | -0.1 | 1.031 | -0.1 |
| | 20 x 20 | 1.053 | 1.047 | -0.6 | 1.055 | 0.2 |
| | 25 x 25 | 1.067 | 1.066 | -0.1 | 1.076 | 0.8 |
| 18 MV | 5 x 5 | 0.901 | 0.908 | 0.8 | 0.909 | 0.9 |
| | 10 x10 | 1 | 1 | - | 1 | - |
| | 15 x15 | 1.045 | 1.049 | 0.5 | 1.044 | -0.1 |
| | 20 x 20 | 1.071 | 1.071 | -0.03 | 1.065 | -0.6 |
| | 25 x 25 | 1.086 | 1.083 | -0.3 | 1.084 | -0.2 |

*Table S2: Electron beam relative output factors normalized to a 10 x 10 cm² field size measured with the Cerenkov and scintillation detector in comparison with a TN30013 Farmer ionization chamber (PTW, Freiburg, Germany) and their relative differences.*

| Energy | Applicator size (cm²) | Ionization chamber | Cerenkov detector | Relative difference (%) | BCF12 detector | Relative difference (%) |
|---|---|---|---|---|---|---|
| 6 MeV | 6 x 6 | 0.964 | 0.964 | 0 | 0.966 | 0.3 |
| | 10 x10 | 1 | 1 | - | 1 | - |
| | 15 x15 | 1.005 | 1.007 | 0.2 | 1.008 | 0.3 |
| | 20 x 20 | 1.012 | 1.010 | -0.2 | 1.010 | -0.2 |
| | 25 x 25 | 1.006 | 1.007 | -0.2 | 1.014 | 0.8 |
| 9 MeV | 6 x 6 | 0.975 | 0.977 | 0.8 | 0.980 | 0.5 |
| | 10 x10 | 1 | 1 | - | 1 | - |
| | 15 x15 | 0.994 | 0.993 | 0.3 | 0.991 | -0.3 |
| | 20 x 20 | 0.983 | 0.975 | -0.8 | 0.980 | -0.3 |
| | 25 x 25 | 0.962 | 0.968 | 0.6 | 0.960 | -0.2 |
| 12 MeV | 6 x 6 | 0.963 | 0.956 | -0.7 | 0.966 | 0.3 |
| | 10 x10 | 1 | 1 | - | 1 | - |
| | 15 x15 | 0.989 | 0.989 | 0 | 0.989 | 0 |
| | 20 x 20 | 0.976 | 0.982 | 0.6 | 0.975 | -0.1 |
| | 25 x 25 | 0.949 | 0.942 | -0.7 | 0.947 | -0.2 |
| 16 MeV | 6 x 6 | 0.987 | 0.984 | -0.3 | 0.993 | 0.6 |
| | 10 x10 | 1 | 1 | - | 1 | - |
| | 15 x15 | 0.982 | 0.984 | 0.2 | 0.983 | 0.1 |
| | 20 x 20 | 0.966 | 0.965 | -0.1 | 0.969 | 0.3 |
| | 25 x 25 | 0.937 | 0.940 | 0.3 | 0.938 | 0.1 |
| 20 MeV | 6 x 6 | 1.006 | 1.004 | -0.2 | 1.009 | 0.3 |
| | 10 x10 | 1 | 1 | - | 1 | - |
| | 15 x15 | 0.973 | 0.966 | -0.6 | 0.971 | -0.2 |
| | 20 x 20 | 0.950 | 0.951 | 0.1 | 0.950 | 0 |
| | 25 x 25 | 0.922 | 0.926 | 0.4 | 0.918 | -0.4 |

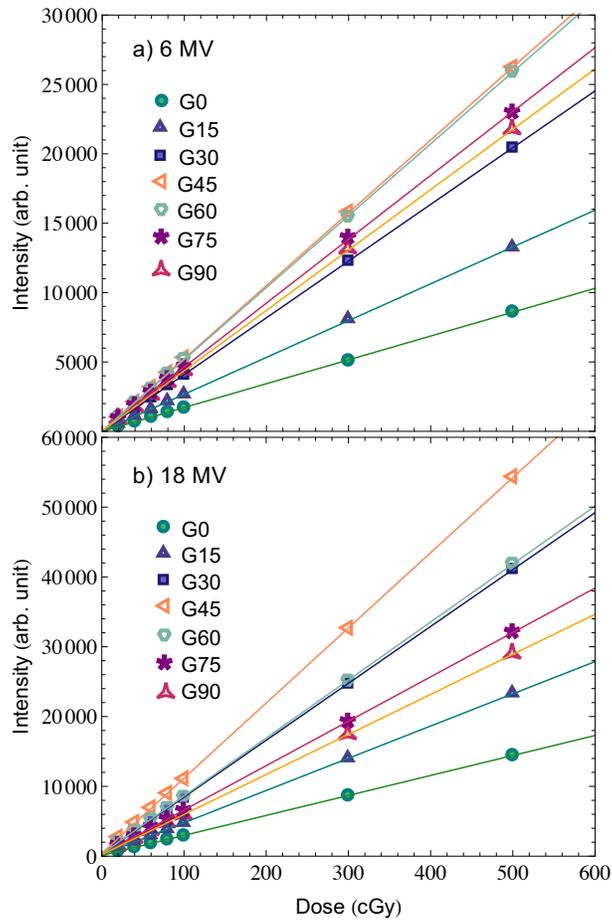

*Figure S3 : Signal emitted by the Cerenkov detector sensitive volume as a function of the dose obtained with (a) 6 MV and (b) 18 MV photon beams at various Gantry angles ranging from 0° to 90°.*

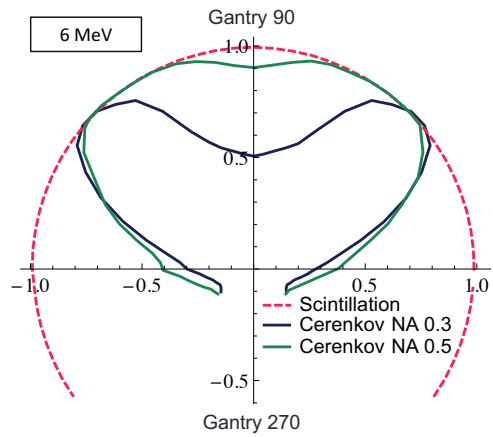

*Figure S4 : Signal emitted by the two Cerenkov detector sensitive volumes (NA=0.3 and NA=0.5) as a function of the angle of incidence for fixed dose using a 6 MeV beam.*